\documentclass[english,onecolumn]{IEEEtran}
\usepackage[T1]{fontenc}
\usepackage[latin9]{inputenc}
\usepackage{array}
\usepackage{float}
\usepackage{multirow}
\usepackage{amsmath}
\usepackage{amsthm}
\usepackage{amssymb}
\usepackage{graphicx}
\usepackage{esint}

\makeatletter

\providecommand{\tabularnewline}{\\}
\floatstyle{ruled}
\newfloat{algorithm}{tbp}{loa}
\providecommand{\algorithmname}{Algorithm}
\floatname{algorithm}{\protect\algorithmname}

\theoremstyle{plain}
\newtheorem{thm}{\protect\theoremname}
\theoremstyle{definition}
\newtheorem{defn}[thm]{\protect\definitionname}
\theoremstyle{plain}
\newtheorem{prop}[thm]{\protect\propositionname}
\theoremstyle{plain}
\newtheorem{cor}[thm]{\protect\corollaryname}
\theoremstyle{plain}
\newtheorem{lem}[thm]{\protect\lemmaname}
\theoremstyle{remark}
\newtheorem{rem}[thm]{\protect\remarkname}

\usepackage{mathrsfs}
\usepackage{algorithm}
\usepackage{algpseudocode}

\usepackage{colortbl}
\definecolor{lightgray}{rgb}{0.9,0.9,0.9}
\definecolor{lightred}{rgb}{1,0.8,0.8}
\definecolor{lightgreen}{rgb}{0.6,1,0.6}
\definecolor{lightyellow}{rgb}{1,1,0.5}
\definecolor{lightgrey}{rgb}{0.8,0.8,0.8}

\allowdisplaybreaks[1]

\usepackage{babel}

\providecommand{\definitionname}{Definition}
\providecommand{\theoremname}{Theorem}

\makeatother

\usepackage{babel}
\providecommand{\corollaryname}{Corollary}
\providecommand{\definitionname}{Definition}
\providecommand{\lemmaname}{Lemma}
\providecommand{\propositionname}{Proposition}
\providecommand{\remarkname}{Remark}
\providecommand{\theoremname}{Theorem}

\begin{document}
\title{Efficient Approximate Minimum Entropy Coupling of Multiple Probability
Distributions}
\author{Cheuk Ting Li\\
 Department of Information Engineering\\
 The Chinese University of Hong Kong\\
 Email: ctli@ie.cuhk.edu.hk}
\maketitle
\begin{abstract}
Given a collection of probability distributions $p_{1},\ldots,p_{m}$,
the minimum entropy coupling is the coupling $X_{1},\ldots,X_{m}$
($X_{i}\sim p_{i}$) with the smallest entropy $H(X_{1},\ldots,X_{m})$.
While this problem is known to be NP-hard, we present an efficient
algorithm for computing a coupling with entropy within 2 bits from
the optimal value. More precisely, we construct a coupling with entropy
within 2 bits from the entropy of the greatest lower bound of $p_{1},\ldots,p_{m}$
with respect to majorization. This construction is also valid when
the collection of distributions is infinite, and when the supports
of the distributions are infinite. Potential applications of our results
include random number generation, entropic causal inference, and functional
representation of random variables.
\end{abstract}


\section{Introduction}

The problem of finding the minimum entropy coupling of two discrete
probability distributions $p,q$, i.e., finding a pair of jointly
distributed random variables $X,Y$ such that $X$ has marginal distribution
$p$, $Y$ has marginal distribution $q$, and the joint entropy $H(X,Y)$
is minimized, has been studied by Vidyasagar \cite{vidyasagar2012metric},
Painsky, Rosset and Feder \cite{painsky2013memoryless,painsky2019innovation},
Kova{\v{c}}evi{\'c}, Stanojevi{\'c} and {\v{S}}enk \cite{kovavcevic2015entropy},
Kocaoglu, Dimakis, Vishwanath and Hassibi \cite{kocaoglu2017entropic,kocaoglu2017greedy},
Cicalese, Gargano and Vaccaro \cite{cicalese2017find,cicalese2019minimum},
Yu and Tan \cite{yu2018asymptotic}, and Rossi \cite{rossi2019greedy}.
Also see \cite{roughgarden2013marginals,han2016mutual,cicalese2016approximating}
for related problems. While it is shown in \cite{vidyasagar2012metric,kovavcevic2015entropy}
that this problem is NP-hard, a polynomial time approximation algorithm
(within 1 bit from the optimum) is given in \cite{cicalese2019minimum}
(also see \cite{kocaoglu2017greedy,rossi2019greedy}).

This problem can be generalized to the coupling of $m$ probability
distributions $p_{1},\ldots,p_{m}$ (i.e., constructing random variables
$X_{1},\ldots,X_{m}$ with marginals $X_{i}\sim p_{i}$), where \cite{cicalese2019minimum}
gives an algorithm for constructing a coupling with entropy $H(X_{1},\ldots,X_{m})$
within $\lceil\log m\rceil$ bits from the optimum (also see \cite{kocaoglu2017greedy}
for another algorithm). More precisely, \cite{cicalese2019minimum}
gives a coupling with entropy at most $H(\bigwedge_{i}p_{i})+\lceil\log m\rceil$
bits, where $\bigwedge_{i}p_{i}$ denotes the greatest lower bound
of $p_{1},\ldots,p_{m}$ with respect to majorization of probability
vectors \cite{marshall1979inequalities}. Since any coupling of $p_{1},\ldots,p_{m}$
has entropy at least $H(\bigwedge_{i}p_{i})$ \cite{cicalese2019minimum},
this gives a construction within $\lceil\log m\rceil$ bits from the
optimum.

In this paper, we improve this result by constructing a coupling of
$p_{1},\ldots,p_{m}$ with entropy at most
\begin{equation}
H\Big(\bigwedge_{i}p_{i}\Big)+2-2^{2-m},\label{eq:main_bd}
\end{equation}
which is at most $2$ bits from the optimum. A more general bound
in terms of R\'enyi entropy \cite{renyi1961measures} can also be
obtained. See Corollary \ref{cor:cp_geom_bd} and Theorem \ref{thm:finite}.
Compared to the $\lceil\log m\rceil$ gap in \cite{cicalese2019minimum},
the gap $2-2^{2-m}\le2$ in our result does not scale with $m$. Also
note that the gap becomes $1$ when $m=2$, the same gap as in \cite{cicalese2019minimum,rossi2019greedy}
for the coupling of two distributions. We describe an algorithm (Algorithm
\ref{alg:finitecp}) for computing a coupling achieving \eqref{eq:main_bd}
with time complexity $O(m^{2}n+mn\log n)$, where we assume the pmf's
$p_{i}$ are over a finite set $\mathcal{X}$ with $|\mathcal{X}|=n$.
If we allow an error at most $\epsilon$ (i.e., changing each $p_{i}$
by at most $\epsilon$ in total variation distance), we can reduce
the time complexity to $O(mn\log(1/\epsilon)+mn\log n)$ (see Remark
\ref{rem:error_cp}).

Moreover, \eqref{eq:main_bd} continues to hold when the collection
of pmf's to be coupled is infinite, or even uncountable (in this case,
$m=\infty$ and $2^{2-m}=0$). The bound in \eqref{eq:main_bd} also
applies to the case where the supports of the distributions are infinite.
These cases are not handled in \cite{kocaoglu2017entropic,kocaoglu2017greedy,cicalese2019minimum,rossi2019greedy}.

Below are some potential applications of a low entropy coupling of
a collection of distributions.

\subsection{Random Number Generation}

It was shown by Knuth and Yao~\cite{knuth1976complexity} that a
discrete random variable $X$ can be generated using an expected number
of fair coin flips no more than $H(X)+2$, indicating that the entropy
$H(X)$ is a measure of the amount of resources (coin flips) needed
to generate the random number $X$ (also see \cite{roche1991,han1997randgen}).
The entropy of a coupling of a collection of distributions $S$ can
be regarded as the amount of resources needed to allow generation
of any distribution in $S$. More precisely, consider the setting
where there is a random number generator device that can output a
random number to the user (who does not have access to random sources
other than the generator). The user wants to generate $X\sim p$ for
a distribution $p\in S$ of the user's choice ($p$ is not fixed a
priori). If the generator is versatile enough to generate any distribution
$p$ at the user's request, then the minimum amount of entropy used
by the generator is $H(p)$. Nevertheless, the generator may not be
programmable or configurable. If we assume the generator is only capable
of generating a random number $Z$ following a fixed distribution
(that depends on the design of the generator, but cannot depend on
the user's choice of $p$), then the user has to apply a mapping $g_{p}$
(depending on the choice of $p$) to obtain the final random number
$X=g_{p}(Z)\sim p$. This induces a coupling $\{g_{p}(Z)\}_{p\in S}$
of the distributions in $S$. Therefore, the minimum entropy coupling
of $S$ corresponds to the distribution of $Z$ that has the minimum
entropy needed to accomplish this task.

Existing hardware random number generators  are capable of generating
a uniformly random integer within a range of integers. While we can
generate from any discrete distribution by repeated and interactive
usages of such generator (e.g. by \cite{knuth1976complexity}), such
interactive communication between the generator and the user may not
be feasible depending on the situation (e.g. delay in generating the
random number and communication). The minimum entropy coupling allows
us to design the generator according to $S$ (with possibly non-uniform
output $Z$) so that we only need to use the generator once per random
number $X\sim p$ obtained by the user.

We will see in the following sections that our construction is similar
to the alias method for random number generation by Walker \cite{walker1977efficient}.
While the alias method only works for discrete distributions with
finite support, and requires an amount of entropy approximately $\log k$
(where $k$ is the size of the support), our construction works for
any discrete distribution (with finite or infinite support), and requires
an amount of entropy close to the theoretical minimum (which can be
much smaller than $\log k$ depending on the collection of distributions
$S$).

A related setting is channel simulation (see \cite{bennett2002entanglement,winter2002compression,cuff2013distributed,bennet2014reverse}
for the asymptotic case, and \cite{harsha2010communication,kumar2014exact,li2018universal}
for the one-shot case), where the encoder observes a distribution
$p\in S$ in a collection of distributions $S$ and transmits a message
$M$ to the decoder (who knows $S$ but does not know $p$ a priori),
so as to allow the decoder to generate $X\sim p$. The aforementioned
random number generation setting corresponds to the one-shot channel
simulation setting where the encoder does not have local randomness,
the communication $M$ from the encoder to the decoder is unlimited,
and our goal is to minimize the amount of local randomness at the
decoder in order to generate $X\sim p$ (we require the distribution
of the local randomness to be fixed).

\subsection{Functional Representation and Entropic Causal Inference}

The functional representation lemma \cite{elgamal2011network} states
that for any pair of random variables $(X,Y)\in\mathcal{X}\times\mathcal{Y}$,
there exists a random variable $Z$ independent of $X$ such that
$Y=g(X,Z)$ is a function of $(X,Z)$. See \cite{hajek1979,willems1985}
for applications of this lemma in information theory. Since $Y_{x}:=g(x,Z)\sim p_{Y|X=x}$,
$\{Y_{x}\}_{x\in\mathcal{X}}$ is a coupling of the conditional distributions
$p_{Y|X=x}$, and hence the problem of finding a functional representation
with the smallest $H(Z)$ is equivalent to the minimum entropy coupling
problem (see \cite{kocaoglu2017entropic,kocaoglu2017greedy}).

Shannon \cite[Fig. 1]{shannon1948mathematical} considers a channel
to be a function mapping the input signal and noise source to the
received signal. Letting the input signal and the received signal
be $X$ and $Y$ respectively, the minimum $H(Z)$ in the functional
representation would be the minimum entropy of the noise source of
the channel. Note that this measure is an inherent property of the
channel, and does not depend on the input distribution $p_{X}$ as
long as $p_{X}(x)>0$ for all $x\in\mathcal{X}$ (since the minimum
$H(Z)$ is the minimum entropy of a coupling of $\{p_{Y|X=x}\}_{x\in\mathcal{X}}$
which does not depend on $p_{X}$).

Kocaoglu, Dimakis, Vishwanath and Hassibi \cite{kocaoglu2017entropic,kocaoglu2017greedy}
consider the problem of identifying the causal direction between $X$
and $Y$, based on the assumption that the correct causal direction
gives a small $H(Z)$. More precisely, the \emph{entropic causal inference}
method declares that $X\to Y$ is the correct direction if $Y=g(X,Z)$
can be achieved with a smaller $H(X)+H(Z)$ compared to the smallest
$H(Y)+H(\tilde{Z})$ satisfying $X=\tilde{g}(Y,\tilde{Z})$. They
have proposed algorithms for minimizing $H(Z)$, or equivalently,
minimizing the entropy of the coupling of $p_{Y|X=x}$ (also see \cite{cicalese2019minimum}
for another algorithm). Nevertheless, these algorithms only work when
$\mathcal{X}$ is finite (or the number of distributions to couple
is finite). The method in this paper works regardless of whether $X$
is a discrete or continuous random variable (though $Y$ must be discrete).
By \eqref{eq:main_bd}, the minimum of $H(Z)$ is closely approximated
by $H(\bigwedge_{x\in\mathcal{X}}p_{Y|X=x})$ (within $2$ bits),
and hence replacing $H(Z)$ by $H(\bigwedge_{x\in\mathcal{X}}p_{Y|X=x})$
(which can be computed in $O(|\mathcal{X}||\mathcal{Y}|\log|\mathcal{Y}|)$
time if $|\mathcal{X}|,|\mathcal{Y}|<\infty$) in the entropic causal
inference method provides a close approximation that can be computed
efficiently (compared to the exact minimization of $H(Z)$ which is
NP-hard \cite{vidyasagar2012metric,kovavcevic2015entropy}). If the
function $g$ is also needed, then it can be computed in $O(|\mathcal{X}|^{2}|\mathcal{Y}|+|\mathcal{X}||\mathcal{Y}|\log|\mathcal{Y}|)$
time using Algorithm \ref{alg:finitecp}.

The problem of minimizing $H(Y|Z)$ (instead of $H(Z)$) was studied
by Li and El Gamal \cite{sfrl_trans}. The strong functional representation
lemma \cite{sfrl_trans} states that for any pair of random variables
$(X,Y)$, there exists a random variable $Z$ independent of $X$
such that $Y$ is a function of $(X,Z)$, and $H(Y|Z)\le I(X;Y)+\log(I(X;Y)+1)+4$
(also see \cite{harsha2010communication,braverman2014public}). The
lemma is applied to show several one-shot variable-length lossy source
coding results, and a short proof of the asymptotic achievability
in the Gelfand-Pinsker theorem \cite{gelfand1980coding}. It is also
used in \cite{li2018minimax} to prove a result on minimax remote
prediction with a communication constraint. The Poisson functional
representation given in \cite{sfrl_trans} (which induces a coupling
of $p_{Y|X=x}$) is also used in \cite{li2019unified} to prove various
results in multi-user information theory. In this paper, we concern
the minimization of $H(Z)$ instead of $H(Y|Z)$ (while \cite{sfrl_trans}
gives a cardinality bound $|\mathcal{Z}|\le|\mathcal{X}|(|\mathcal{Y}|-1)+2$
in addition to the bound on $H(Y|Z)$, this is not the main objective
there).

\subsection{Other Uses of Coupling of Collections of Distributions}

It has been shown that for any collection of distributions $p_{1},\ldots,p_{m}$,
it is possible to find a coupling $X_{1},\ldots,X_{m}$ such that
\begin{equation}
\mathbf{P}(X_{i}\neq X_{j})\le2d_{\mathrm{TV}}(p_{i},p_{j})\label{eq:tv_bd}
\end{equation}
for any $i,j$, where $d_{\mathrm{TV}}$ is the total variation distance.
This was shown in \cite{broder1997resemblance} for uniform distributions,
\cite{kleinberg2002approximation} for discrete distributions, and
\cite{angel2019pairwise,li2019pairwise} for general distributions.
This result was used in locality sensitive hashing \cite{charikar2002similarity}
and randomized rounding algorithms \cite{kleinberg2002approximation,barak2008rounding}.
While a coupling achieving \eqref{eq:tv_bd} is likely to have low
entropy (since many values of $X_{i}$ are the same), a low entropy
coupling does not necessarily have a low $\mathbf{P}(X_{i}\neq X_{j})$
(since whether $X_{i}\neq X_{j}$ is irrelevant in the calculation
of entropy). We also remark that the connection between entropy and
total variation distance has been studied in \cite{sason2013entropy}
using coupling.

In the study of Markov chains, it is often useful to represent the
Markov chain $X_{1},X_{2},\ldots$ using the functional representation
$X_{n}=g(X_{n-1},Z_{n})$, where $Z_{n}\stackrel{iid}{\sim}p_{Z}$.
In the coupling from the past algorithm for sampling from the stationary
distribution of a Markov chain \cite{propp1996exact,propp1998coupling},
the function $g$ is designed so that $g(x,z)$ are likely to be equal
for different values of $x$. This representation is also referred
as innovation representation in \cite{painsky2019innovation}. Since
the minimum entropy of $Z_{n}$ is the entropy of the minimum entropy
coupling of $\{p_{X_{n}|X_{n-1}=x}\}_{x}$, we can apply the coupling
achieving \eqref{eq:main_bd} in this paper to generate $X_{1},X_{2},\ldots$
using a small entropy rate of $Z_{1},Z_{2},\ldots$.

Other works on the coupling of a collection of distributions include
Wasserstein barycenter \cite{agueh2011barycenters} and multi-marginal
optimal transport \cite{kellerer1984duality,gangbo1998optimal,pass2011uniqueness,li2019pairwise}.

\subsection*{Notations}

Throughout this paper, we assume that $\log$ is to base $2$ and
the entropy $H$ is in bits. We write $\mathbb{N}=\{1,2,3,\ldots\}$,
$[n]=\{1,\ldots,n\}$. For a statement $E$, we write $\mathbf{1}\{E\}$
for the indicator function where $\mathbf{1}\{E\}=1$ if $E$ holds,
$\mathbf{1}\{E\}=0$ otherwise.

A right stochastic matrix is a square matrix with nonnegative entries
where each row sums to $1$. The support of a probability mass function
(pmf) $p$ is written as $\mathrm{supp}(p)$. For a pmf $p$ over
$[n]$, its probability vector is denoted as $\vec{p}\in\mathbb{R}^{n}$
(a row vector). For a pmf $p$ over the set $\mathcal{X}$, and a
pmf $q$ over the set $\mathcal{Y}$, the product pmf $p\times q$
is a pmf over $\mathcal{X}\times\mathcal{Y}$ with $(p\times q)(x,y):=p(x)q(y)$.
The pmf of the Bernoulli distribution is denoted as $\mathrm{Bern}_{\gamma}(x):=\mathbf{1}\{x=0\}(1-\gamma)+\mathbf{1}\{x=1\}\gamma$.
The pmf of the geometric distribution over $\mathbb{N}$ is denoted
as $\mathrm{Geom}_{\gamma}(x):=\gamma(1-\gamma)^{x-1}$. The pmf of
the capped geometric distribution over $[k]$ is denoted as 
\begin{equation}
\mathrm{CGeom}_{\gamma,k}(x):=\begin{cases}
\gamma(1-\gamma)^{x-1} & \mathrm{if}\;x<k\\
(1-\gamma)^{k-1} & \mathrm{if}\;x=k\\
0 & \mathrm{if}\;x>k.
\end{cases}\label{eq:cgeom}
\end{equation}
The R\'enyi entropy \cite{renyi1961measures} of a pmf $p$ is defined
as
\[
H_{\alpha}(p):=\frac{1}{1-\alpha}\log\sum_{x\in\mathrm{supp}(p)}(p(x))^{\alpha}
\]
for $\alpha\in\mathbb{R}_{\ge0}\backslash\{1\}$. When $\alpha=1$,
$H_{\alpha}(p):=H(p)$ is the Shannon entropy. When $\alpha=\infty$,
$H_{\alpha}(p):=-\log\max_{x}p(x)$.

\medskip{}

\section{Coupling and Majorization\label{sec:mrq}}

We first define a coupling of a set of distributions.
\begin{defn}
For a set of pmf's $S$, we say that an indexed set of random variables
$\{X_{p}\}_{p\in S}$ is a coupling of $S$, written as $\{X_{p}\}_{p\in S}\in\Gamma(S)$,
if $X_{p}$ has marginal distribution $p$ for any $p\in S$. We say
that a pmf $q$ is an \emph{underlying distribution of a coupling}
of $S$, written as $q\in\tilde{\Gamma}(S)$, if there exists $\{X_{p}\}_{p\in S}\in\Gamma(S)$
and random variable $Z\sim q$ such that $X_{p}$ is a function of
$Z$ for all $p\in S$.\footnote{Technically, to make the set $\tilde{\Gamma}(S)$ well-defined, we
can restrict $q$ to be a pmf over $\mathbb{N}$, which will not cause
any loss of generality since the support of a pmf is always countable.
}

Define the \emph{minimum R\'enyi entropy of couplings} of a set of
pmf's $S$ as
\begin{equation}
H_{\alpha}^{*}(S):=\inf_{q\in\tilde{\Gamma}(S)}H_{\alpha}(q)\label{eq:mrec}
\end{equation}
for $\alpha\in\mathbb{R}_{\ge0}\cup\{\infty\}$. We write $H^{*}(S)=H_{1}^{*}(S)$.
It is straightforward to show that when $S=\{p_{1},\ldots,p_{m}\}$
is finite, then $H_{\alpha}^{*}(S)=\inf_{\{X_{p_{i}}\}_{i\in[m]}\in\Gamma(S)}H_{\alpha}(X_{p_{1}},\ldots,X_{p_{m}})$
(to show a one-to-one correspondence between $\tilde{\Gamma}(S)$
and $\Gamma(S)$, simply take $q\in\tilde{\Gamma}(S)$ to be the joint
pmf of $\{X_{p_{i}}\}_{i\in[m]}$). Nevertheless, we define $H_{\alpha}^{*}(S)$
in \eqref{eq:mrec} for general $S$ using $\tilde{\Gamma}(S)$ instead
of $\Gamma(S)$, in order to avoid working with the joint entropy
of an infinite collection of random variables when $|S|=\infty$.
\end{defn}
The goal of this paper is to find or approximate $H_{\alpha}^{*}(S)$.
We present the concept of aggregation in \cite{vidyasagar2012metric,cicalese2016approximating}.
\begin{defn}
For two pmf's $p,q$, we say $p$ is an \emph{aggregation} of $q$,
written as $q\sqsubseteq p$, if there exists a function $g:\mathrm{supp}(q)\to\mathrm{supp}(p)$
(called the \emph{aggregation map}) such that $p$ is the pmf of $g(X)$,
where $X\sim q$.
\end{defn}
It is clear that ``$\sqsubseteq$'' is a transitive relation. If
$p,q$ are pmf's over $[n]$, then $q\sqsubseteq p$ if and only if
there exists a right stochastic matrix $M$ with $\{0,1\}$ entries
such that the probability vectors satisfy $\vec{p}=\vec{q}M$. Note
that $q\in\tilde{\Gamma}(S)$ if and only if $q\sqsubseteq p$ for
any $p\in S$. Therefore, a coupling of $S$ can be specified using
an underlying distribution $q\in\tilde{\Gamma}(S)$ and the set of
aggregation maps $\{g_{p}\}_{p\in S}$, where $g_{p}$ is the aggregation
map for $q\sqsubseteq p$.

We will then show that ``$\sqsubseteq$'' is ``closed under pointwise
limit'' in the following sense:
\begin{prop}
\label{prop:agg_closed}Let $q$ be a pmf over a countable set $\mathcal{X}$,
and $p,p_{1},p_{2},\ldots$ be pmf's over a countable set $\mathcal{Y}$,
such that $p(y)=\lim_{i\to\infty}p_{i}(y)$ for any $y\in\mathcal{Y}$,
and $q\sqsubseteq p_{i}$ for any $i\ge1$, then we have $q\sqsubseteq p$.
\end{prop}
\begin{IEEEproof}
Without loss of generality, assume $\mathcal{X}=\mathcal{Y}=\mathbb{N}$,
and $q(1)\ge q(2)\ge\cdots$. Let $g_{1},g_{2},\ldots$ be functions
from $\mathbb{N}$ to $\mathbb{N}$ such that $g_{i}(X)\sim p_{i}$,
where $X\sim q$. Consider whether $g_{i}(1)=1$. There exists an
increasing sequence $i_{1},i_{2},\ldots$ such that $\mathbf{1}\{g_{i_{j}}(1)=1\}$
are the same for all $j$ (since there are only two possibilities
of $\mathbf{1}\{g_{i_{j}}(1)=1\}\in\{0,1\}$). Let that value of $\mathbf{1}\{g_{i_{j}}(1)=1\}$
be $b_{1,1}$. There exists an increasing subsequence $i'_{1},i'_{2},\ldots$
of $i_{1},i_{2},\ldots$ such that for any $x,y\le2$, $\mathbf{1}\{g_{i'_{j}}(x)=y\}$
are the same for all $j$ (since there are only $2^{4}$ possibilities
of $\{\mathbf{1}\{g_{i'_{j}}(x)=y\}\}_{x,y\le2}$). Let those values
of $\mathbf{1}\{g_{i'_{j}}(x)=y\}$ be $b_{x,y}$ for $x,y\le2$.
Repeat this procedure to define $b_{x,y}$ for any $x,y\in\mathbb{N}$.

Define $g:\mathrm{supp}(q)\to\mathbb{N}$ by $g(x)=y$ if $b_{x,y}=1$.
We now check that $g$ is well-defined and $g(X)\sim p$. It is clear
from the definition that there does not exist $x$ and $y\neq y'$
such that $b_{x,y}=b_{x,y'}=1$ (consider the $\max\{x,y,y'\}$-th
iteration of the above procedure). Fix any $a,y\in\mathbb{N}$ and
$\epsilon>0$, and consider the $\max\{a,y\}$-th iteration of the
above procedure that fixes $b_{x,y}$ for $x\le a$. There exists
an increasing sequence $i_{1},i_{2},\ldots$ such that $\mathbf{1}\{g_{i_{j}}(x)=y\}=b_{x,y}$
for all $j$ and $x\le a$. By $p(y)=\lim_{i\to\infty}p_{i}(y)$,
there exists $j$ such that $|p(y)-p_{i_{j}}(y)|\le\epsilon$. Since
$p_{i_{j}}(y)=\sum_{x}\mathbf{1}\{g_{i_{j}}(x)=y\}q(x)$, we have
\[
\Big|p(y)-\sum_{x\le a}b_{x,y}q(x)\Big|\le\epsilon+\sum_{x>a}q(x).
\]
Taking $a\to\infty$ and $\epsilon\to0$, we have $\sum_{x}b_{x,y}q(x)=p(y)$.
Since $\sum_{x}\sum_{y}b_{x,y}q(x)=\sum_{y}p(y)=1$, for any $x\in\mathrm{supp}(q)$
(where $q(x)>0$), there exists at least one (and thus exactly one)
$y$ such that $b_{x,y}=1$. The result follows.
\end{IEEEproof}
It is demonstrated in \cite{cicalese2019minimum} that majorization
is a useful tool in the study of coupling. Here we present the concept
of majorization which allows infinite sequences or pmf's with infinite
support (e.g. see \cite{marshall1979inequalities}):
\begin{defn}
For two pmf's $p,q$, we say $q$ is \emph{majorized by} $p$, written
as $q\preceq p$, if 
\[
\max_{B\subseteq\mathrm{supp}(q):\,|B|\le k}q(B)\le\max_{A\subseteq\mathrm{supp}(p):\,|A|\le k}p(A)
\]
for any $k\in\mathbb{N}$, where we write $p(A):=\sum_{x\in A}p(x)$.
In other words, the sum of the $k$ largest $q(x)$'s is not greater
than the sum of the $k$ largest $p(x)$'s.
\end{defn}
It is clear that ``$\preceq$'' is a transitive relation. It is
shown in \cite{cicalese2016approximating} that $q\sqsubseteq p$
implies $q\preceq p$. If $p,q$ are pmf's over $[n]$, then it has
been shown that $q\preceq p$ if and only if there exists a doubly
stochastic matrix (i.e., square matrix with nonnegative entries where
each row and column sums to $1$) $M$ such that the probability vectors
satisfy $\vec{p}=\vec{q}M$ (e.g. see \cite{marshall1979inequalities}).
Also, if $p,q$ are pmf's over $[n]$ sorted in descending order (i.e.,
$p(1)\ge p(2)\ge\cdots\ge p(n)$ and likewise for $q$), then $q\preceq p$
if and only if there exists a lower triangular right stochastic matrix
$M$ such that $\vec{p}=\vec{q}M$. This property will be strenghtened
in Lemma \ref{lem:alias}.

Also note that R\'enyi entropy is Schur concave \cite{marshall1979inequalities},
i.e., we have $H_{\alpha}(q)\ge H_{\alpha}(p)$ if $q\preceq p$.
We then prove a useful property of majorization and aggregation:
\begin{prop}
\label{prop:kernel_maj}Let $X$ be a random variable with pmf $p_{X}$,
and $Y$ be a random variable with conditional pmf $p_{Y|X=x}$, and
$p_{X,Y}$ be the joint pmf of $(X,Y)$. Define $\tilde{X},p_{\tilde{X}},\tilde{Y},p_{\tilde{Y}|\tilde{X}=x},p_{\tilde{X},\tilde{Y}}$
similarly. We have:
\begin{itemize}
\item If $p_{X}=p_{\tilde{X}}$ and $p_{Y|X=x}\sqsubseteq p_{\tilde{Y}|\tilde{X}=x}$
for all $x$, then $p_{X,Y}\sqsubseteq p_{\tilde{X},\tilde{Y}}$.
\item If $p_{X}=p_{\tilde{X}}$ and $p_{Y|X=x}\preceq p_{\tilde{Y}|\tilde{X}=x}$
for all $x$, then $p_{X,Y}\preceq p_{\tilde{X},\tilde{Y}}$.
\end{itemize}
\end{prop}
\begin{IEEEproof}
Assume $p_{X}=p_{\tilde{X}}$ and $p_{Y|X=x}\sqsubseteq p_{\tilde{Y}|\tilde{X}=x}$
for all $x$. There exists functions $g_{x}$ for $x\in\mathrm{supp}(p_{X})$
such that $(X,g_{X}(Y))\stackrel{d}{=}(\tilde{X},\tilde{Y})$. Hence,
$p_{X,Y}\sqsubseteq p_{\tilde{X},\tilde{Y}}$ with the aggregation
map $(x,y)\mapsto(x,g_{x}(y))$.

Assume $p_{X}=p_{\tilde{X}}$ and $p_{Y|X=x}\preceq p_{\tilde{Y}|\tilde{X}=x}$
for all $x$. Fix any $A\subseteq\mathrm{supp}(p_{X,Y})$. For any
$x$, let $B_{x}$ attains the maximum in $\max_{B\subseteq\mathrm{supp}(p_{\tilde{Y}|\tilde{X}=x}):\,|B|\le|\{y:\,(x,y)\in A\}|}q(B)$.
Since $p_{Y|X=x}\preceq p_{\tilde{Y}|\tilde{X}=x}$, we have 
\begin{align*}
p_{X,Y}(A) & =\sum_{x}p_{X}(x)\sum_{y:\,(x,y)\in A}p_{Y|X=x}(y)\\
 & \le\sum_{x}p_{X}(x)p_{\tilde{Y}|\tilde{X}=x}(B_{x})\\
 & =p_{\tilde{X},\tilde{Y}}\left(\left\{ (x,y):\,y\in B_{x}\right\} \right).
\end{align*}
Since $|\{(x,y):\,y\in B_{x}\}|\le|A|$, we have $p_{X,Y}\preceq p_{\tilde{X},\tilde{Y}}$.
\end{IEEEproof}
We then present the definition of the greatest lower bound (see e.g.
\cite{cicalese2002supermodularity} for the finite case):
\begin{defn}
For a set of pmf's $S$ where 
\begin{equation}
\lim_{k\to\infty}\inf_{p\in S}\max_{A\subseteq\mathrm{supp}(p):\,|A|\le k}p(A)=1,\label{eq:glb_cond}
\end{equation}
define its \emph{greatest lower bound with respect to majorization},
written as $q=\bigwedge S$, as a pmf $q$ over $\mathbb{N}$ given
by
\[
q(k):=\inf_{p\in S}\max_{A\subseteq\mathrm{supp}(p):\,|A|\le k}p(A)-\inf_{p\in S}\max_{A\subseteq\mathrm{supp}(p):\,|A|\le k-1}p(A).
\]
If \eqref{eq:glb_cond} is not satisfied, then $\bigwedge S$ is undefined.
(Note that \eqref{eq:glb_cond} is always satisfied when $S$ is finite.)
\end{defn}
If $S$ contains pmf's over the set $\mathcal{X}$, and $|S|=m<\infty$,
$|\mathcal{X}|=n<\infty$, then it is clear that $\bigwedge S$ can
be computed in $O(mn\log n)$ time (by sorting the pmf's and computing
partial sums). We give some properties of the greatest lower bound.
Note that the case $|S|=2$ has been shown in \cite{cicalese2002supermodularity}.
While it is straightforward to generalize it to $|S|>2$ and $|S|=\infty$,
we state these properties for the sake of completeness. 
\begin{prop}
For a set of pmf's $S$, if $q=\bigwedge S$ exists, then
\begin{enumerate}
\item $q(1)\ge q(2)\ge q(3)\ge\cdots$.
\item $q\preceq p$ for any $p\in S$.
\item For any $\tilde{q}$ such that $\tilde{q}\preceq p$ for any $p\in S$,
we have $\tilde{q}\preceq q$.
\end{enumerate}
\end{prop}
\begin{IEEEproof}
Note that $q(1)\ge q(2)\ge\cdots$ is equivalent to the concavity
of $\inf_{p\in S}\max_{A\subseteq\mathrm{supp}(p):\,|A|\le k}p(A)$
in $k$, which holds because the infimum of concave functions is concave.
We have $\sum_{i=1}^{k}q(i)=\inf_{p\in S}\max_{A\subseteq\mathrm{supp}(p):\,|A|\le k}p(A)$,
and hence $q\preceq p$ for any $p\in S$. For any $\tilde{q}$ such
that $\tilde{q}\preceq p$ for any $p\in S$, we have $\max_{A\subseteq\mathrm{supp}(\tilde{q}):\,|A|\le k}\tilde{q}(A)\le\inf_{p\in S}\max_{A\subseteq\mathrm{supp}(p):\,|A|\le k}p(A)=\sum_{i=1}^{k}q(i)$,
and hence $\tilde{q}\preceq q$.
\end{IEEEproof}
As a result of these properties, if $\bigwedge S$ exists, for any
$q\in\tilde{\Gamma}(S)$, we have $q\preceq\bigwedge S$, and hence
$H_{\alpha}(q)\ge H_{\alpha}(\bigwedge S)$. Therefore, $H_{\alpha}^{*}(S)\ge H_{\alpha}(\bigwedge S)$.

\medskip{}

\section{Coupling by Geometric Splitting\label{sec:geomsplit}}

We now present the main result in this paper, which shows that if
the pmf's $p,q$ satisfy $q\preceq p$, then after splitting each
mass $q(y)$ into a sequence of masses $q(y)/2$, $q(y)/4$, $q(y)/8$,...
(or equivalently, consider the joint pmf of $(Y,Z)$ where $Y\sim q$
is independent of $Z\sim\mathrm{Geom}_{1/2}$), then $p$ will be
an aggregation of the resultant pmf $q\times\mathrm{Geom}_{1/2}$
(``$\times$'' denote the independent product of two pmf's, i.e.,
it is the pmf of $(Y,Z)$ mentioned before; refer to the notation
section for the definition), which we call the \emph{geometric splitting}
of $q$. 
\begin{thm}
\label{thm:geomsplit}If $q\preceq p$, then 
\[
q\times\mathrm{Geom}_{1/2}\sqsubseteq p.
\]
\end{thm}
\medskip{}

A direct result of this theorem is the following explicit formula
of an underlying distribution of a coupling.
\begin{cor}
\label{cor:cp_geom_bd}For a set of pmf's $S$, if $\bigwedge S$
exists, then
\[
\Big(\bigwedge S\Big)\times\mathrm{Geom}_{1/2}\in\tilde{\Gamma}(S).
\]
As a result, the minimum R\'enyi entropy of couplings of $S$ satisfies
\[
H_{\alpha}\Big(\bigwedge S\Big)\le H_{\alpha}^{*}(S)\le H_{\alpha}\Big(\bigwedge S\Big)+H_{\alpha}(\mathrm{Geom}_{1/2}),
\]
where
\[
H_{\alpha}(\mathrm{Geom}_{1/2})=\begin{cases}
\infty & \mathrm{if}\;\alpha=0\\
2 & \mathrm{if}\;\alpha=1\\
1 & \mathrm{if}\;\alpha=\infty\\
\frac{-\alpha-\log(1-2^{-\alpha})}{1-\alpha} & \mathrm{otherwise}
\end{cases}
\]
is the R\'enyi entropy of $\mathrm{Geom}_{1/2}$.
\end{cor}
\medskip{}

Another way to state Theorem \ref{thm:geomsplit} is that for any
pmf $q$, we have $q\times\mathrm{Geom}_{1/2}\in\tilde{\Gamma}(\{p\;\text{pmf over}\;\mathbb{N}:\,q\preceq p\})$.

Before we prove Theorem \ref{thm:geomsplit}, we present a lemma similar
to the alias method \cite{walker1977efficient}, and is a special
case of the algorithm in \cite{cicalese2019minimum}. We include a
proof of the claim for the sake of completeness, and describe a linear
time algorithm (Algorithm \ref{alg:alias}) which is considerably
simpler than that in \cite{cicalese2019minimum}.
\begin{lem}
\label{lem:alias}For any pmf's $p,q$ over $[n]$ such that $q\preceq p$,
$p(1)\ge p(2)\ge\cdots\ge p(n)$ and $q(1)\ge q(2)\ge\cdots\ge q(n)$,
there exists $a_{x}\in[x-1]$ and $0\le r_{x}\le q(x)$ for $x=2,\ldots,n$
such that 
\begin{equation}
p(x)=q(x)-r_{x}+\sum_{y:\,a_{y}=x}r_{y}\label{eq:stoc_tr}
\end{equation}
for any $x=1,\ldots,n$ (we let $r_{1}=0$). Moreover, $a_{x},r_{x}$
can be computed in $O(n)$ time (see Algorithm \ref{alg:alias}).
\end{lem}
Lemma \ref{lem:alias} can be stated in the following more compact
form using matrices. For any pmf's $p,q$ over $[n]$ sorted in descending
order such that $q\preceq p$, there exists a lower triangular right
stochastic matrix $M$ where each row has at most one positive off-diagonal
entry, and the probability vectors satisfy $\vec{p}=\vec{q}M$. Its
equivalence to Lemma \ref{lem:alias} can be shown by letting $M_{x,a_{x}}=r_{x}/q(x)$
and $M_{x,x}=1-r_{x}/q(x)$ for $x\in[n]$ (all other entries of $M$
are zeros).
\begin{IEEEproof}
[Proof of Lemma \ref{lem:alias}]Let $B_{x}:=\{y\in[n]:\,a_{y}=x\}$.
We give $B_{x}$ and $r_{x}$ recursively. Take $B_{n}=\emptyset$,
$r_{n}=q(n)-p(n)$ ($r_{n}\ge0$ since $q\preceq p$). Assume $B_{x+1},\ldots,B_{n}$
and $r_{x+1},\ldots,r_{n}$ are defined and satisfies that $B_{x+1},\ldots,B_{n}$
are disjoint, $B_{x'}\subseteq\{x'+1,\ldots,n\}$ and
\begin{equation}
p(x')=q(x')-r_{x'}+\sum_{y\in B_{x'}}r_{y}\label{eq:ind_hyp}
\end{equation}
for all $x'>x$. We now define $B_{x},r_{x}$. Take
\[
B_{x}=\{t,\ldots,n\}\backslash\bigcup_{y=x+1}^{n}B_{y},
\]
where $t\in\{x+1,\ldots,n+1\}$ such that $\sum_{y\in B_{x}}r_{y}\in[p(x)-q(x),\,p(x)]$.
Such $t$ exists since $r_{y}\le q(y)\le q(x)$ for $y>x$, and
\begin{align*}
 & \sum_{y\in\{x+1,\ldots,n\}\backslash\bigcup_{y'=x+1}^{n}B_{y'}}r_{y}\\
 & =\sum_{y=x+1}^{n}r_{y}-\sum_{y'=x+1}^{n}\sum_{y\in B_{y'}}r_{y}\\
 & \stackrel{(a)}{=}\sum_{y=x+1}^{n}r_{y}-\sum_{y=x+1}^{n}\left(p(y)-q(y)+r_{y}\right)\\
 & =\sum_{y=x+1}^{n}\left(q(y)-p(y)\right)\\
 & =p(x)-q(x)+\sum_{y=x}^{n}\left(q(y)-p(y)\right)\\
 & \ge p(x)-q(x)
\end{align*}
since $q\preceq p$, where (a) is by \eqref{eq:ind_hyp}, and hence
when $t$ decreases from $n+1$ to $x+1$, $\sum_{y\in B_{x}}r_{y}$
increases from $0$ to $\ge p(x)-q(x)$, with step size at most $q(x)$,
and thus there exists $t$ such that $\sum_{y\in B_{x}}r_{y}\in[p(x)-q(x),\,p(x)]$.
In practice, to find $B_{x}$, we only need to scan the elements in
$\bar{B}_{x}:=[n]\backslash\bigcup_{y=x+1}^{n}B_{y}$ in decreasing
order, and add elements from $\bar{B}_{x}$ to $B_{x}$ until $\sum_{y\in B_{x}}r_{y}\ge p(x)-q(x)$.
We then take
\[
r_{x}=q(x)-p(x)+\sum_{y\in B_{x}}r_{y}.
\]
Therefore, we have defined $B_{x},r_{x}$ (and hence $a_{x}$) recursively.

For the running time complexity, note that since $\bar{B}_{x}$ is
decreasing as $x$ decreases, only the elements in $\bar{B}_{x}\backslash\bar{B}_{x-1}$
are relevant to the computation of $B_{x},r_{x}$. Since each $y\in[n]$
can only be removed from $\bar{B}_{x}$ once (i.e., $y\in\bar{B}_{x}\backslash\bar{B}_{x-1}$
for at most one $x$), the overall time complexity is $O(n)$. Also
note that the $B_{x}$ produced by this method must be contiguous
segments of integers, and $\bar{B}_{x}$ must be in the form $\{1,\ldots,|\bar{B}_{x}|\}$,
which allows simpler implementations (e.g. we only need to store $b_{x}:=|\bar{B}_{x}|$
instead of $\bar{B}_{x}$). Refer to Algorithm \ref{alg:alias} (which
we call the \emph{majorized alias} algorithm) for the precise description.
\end{IEEEproof}
\begin{algorithm}[H]
\textbf{$\;\;\;\;$Input:} pmf's $p,q$ over $[n]$ such that $q\preceq p$,
$p(1)\ge\cdots\ge p(n)$ and $q(1)\ge\cdots\ge q(n)$

\textbf{$\;\;\;\;$Output:} $a_{x},r_{x}$ for $x\in\{2,\ldots,n\}$

\smallskip{}

\begin{algorithmic}

\State{$b\leftarrow n$}

\State{$a_{x}\leftarrow1$ for $x=2,\ldots,n$}

\For{$x\leftarrow n,n-1,\ldots,2$}

\State{$r_{x}\leftarrow q(x)-p(x)$}

\While{$r_{x}<0$}

\State{$r_{x}\leftarrow r_{x}+r_{b}$}

\State{$a_{b}\leftarrow x$}

\State{$b\leftarrow b-1$}

\EndWhile

\EndFor

\Return{$\{a_{x}\},\{r_{x}\}$}

\end{algorithmic}

\caption{\label{alg:alias}$\textsc{MajorizedAlias}(p,q)$}
\end{algorithm}

Algorithm \ref{alg:alias} has time complexity $O(n)$ since the block
inside the while loop is executed at most $n$ times ($b$ decreases
each time it is executed). We remark that Algorithm \ref{alg:alias}
reduces to the alias method \cite{walker1977efficient} when $q$
is the uniform distribution. The alias method is an efficient algorithm
that can generate a random number following the distribution $p$
over $[n]$, using a uniformly random integer in $[n]$ and a uniformly
random real number in $[0,1]$. The alias method requires an $O(n)$
(or $O(n\log n)$ if $p$ is unsorted and needs to be sorted first)
precomputation time to compute $a_{x}$ and $r_{x}$ satisfying \eqref{eq:stoc_tr}
(where $q$ is the uniform distribution over $[n]$). After the precomputation,
each sample of $x\sim p$ can be generated in constant time by first
generating $y\sim q$ independent of $z\sim\mathrm{Unif}[0,1]$, and
then outputting $x=y$ if $z\ge r_{y}/q(y)$, $x=a_{y}$ if $z<r_{y}/q(y)$.
While the alias method focuses only on the case where $q$ is uniform
(which guarantees $q\preceq p$), Algorithm \ref{alg:alias} generalizes
it to any $q$ satisfying $q\preceq p$.

Table \ref{tab:alias_eg} shows Algorithm \ref{alg:alias} applied
on $\vec{p}=[0.37,0.36,0.25,0.02,0]$, $\vec{q}=[0.3,0.3,0.2,0.1,0.1]$.
The values of $\{r_{y}\},\{a_{y}\}$ for each iteration $x=5,4,3,2,1$
in the algorithm are given. The red cells are cells with positions
$y$ in the interval $y\in[x..b]$, which are unfinished cells with
$r_{y}$ (the excess amount) computed, but $a_{y}$ is not computed
yet, i.e., it is not yet known where the excess amount $r_{y}$ will
be allocated (while Algorithm \ref{alg:alias} initializes $a_{y}$
to $1$, here we assume $a_{y}$ is initialized to be undefined for
the sake of clarity). The green cells are cells $y$ in the interval
$y\in[b+1\,..\,n]$, which are finished cells with $r_{y},a_{y}$
computed. At each iteration $x$, we keep allocating the the excess
amount of the right-most red cell to the current cell $x$ (and change
the right-most red cell to green), until $q(x)$ plus the total excess
amount allocated to the current cell is at least $p(x)$. The amount
in excess ($q(x)$ plus the total excess amount allocated to the current
cell minus $p(x)$) is written to the $r_{x}$ of the current cell.

\begin{table}
\begin{centering}
\begin{tabular}{|c|c|c|c|c|c|c|}
\cline{2-7} \cline{3-7} \cline{4-7} \cline{5-7} \cline{6-7} \cline{7-7} 
\multicolumn{1}{c|}{} & $y$ & 1 & 2 & 3 & 4 & 5\tabularnewline
\cline{2-7} \cline{3-7} \cline{4-7} \cline{5-7} \cline{6-7} \cline{7-7} 
\multicolumn{1}{c|}{} & $p(y)$ & 0.37 & 0.36 & 0.25 & 0.02 & 0\tabularnewline
\cline{2-7} \cline{3-7} \cline{4-7} \cline{5-7} \cline{6-7} \cline{7-7} 
\multicolumn{1}{c|}{} & $q(y)$ & 0.3 & 0.3 & 0.2 & 0.1 & 0.1\tabularnewline
\hline 
\multirow{2}{*}{Step $x=5$} & $r_{y}$ &  &  &  &  & \cellcolor{lightred}0.1\tabularnewline
\cline{2-7} \cline{3-7} \cline{4-7} \cline{5-7} \cline{6-7} \cline{7-7} 
 & $a_{y}$ &  &  &  &  & \cellcolor{lightred}\tabularnewline
\hline 
\hline 
\multirow{2}{*}{Step $x=4$} & $r_{y}$ &  &  &  & \cellcolor{lightred}0.08 & \cellcolor{lightred}0.1\tabularnewline
\cline{2-7} \cline{3-7} \cline{4-7} \cline{5-7} \cline{6-7} \cline{7-7} 
 & $a_{y}$ &  &  &  & \cellcolor{lightred} & \cellcolor{lightred}\tabularnewline
\hline 
\hline 
\multirow{2}{*}{Step $x=3$} & $r_{y}$ &  &  & \cellcolor{lightred}0.05 & \cellcolor{lightred}0.08 & \cellcolor{lightgreen}0.1\tabularnewline
\cline{2-7} \cline{3-7} \cline{4-7} \cline{5-7} \cline{6-7} \cline{7-7} 
 & $a_{y}$ &  &  & \cellcolor{lightred} & \cellcolor{lightred} & \cellcolor{lightgreen}3\tabularnewline
\hline 
\hline 
\multirow{2}{*}{Step $x=2$} & $r_{y}$ &  & \cellcolor{lightred}0.02 & \cellcolor{lightred}0.05 & \cellcolor{lightgreen}0.08 & \cellcolor{lightgreen}0.1\tabularnewline
\cline{2-7} \cline{3-7} \cline{4-7} \cline{5-7} \cline{6-7} \cline{7-7} 
 & $a_{y}$ &  & \cellcolor{lightred} & \cellcolor{lightred} & \cellcolor{lightgreen}2 & \cellcolor{lightgreen}3\tabularnewline
\hline 
\hline 
\multirow{2}{*}{Step $x=1$} & $r_{y}$ & \cellcolor{lightred}0 & \cellcolor{lightgreen}0.02 & \cellcolor{lightgreen}0.05 & \cellcolor{lightgreen}0.08 & \cellcolor{lightgreen}0.1\tabularnewline
\cline{2-7} \cline{3-7} \cline{4-7} \cline{5-7} \cline{6-7} \cline{7-7} 
 & $a_{y}$ & \cellcolor{lightred} & \cellcolor{lightgreen}1 & \cellcolor{lightgreen}1 & \cellcolor{lightgreen}2 & \cellcolor{lightgreen}3\tabularnewline
\hline 
\end{tabular}
\par\end{centering}
\medskip{}

\caption{\label{tab:alias_eg}Algorithm \ref{alg:alias} applied on $\vec{p}=[0.37,0.36,0.25,0.02,0]$,
$\vec{q}=[0.3,0.3,0.2,0.1,0.1]$.}

\end{table}

\[
\]
We now give a sketch of the proof of Theorem \ref{thm:geomsplit}.
Let $q\preceq p$ with $p(1)\ge p(2)\ge\cdots$ and $q(1)\ge q(2)\ge\cdots$.
Assume they have finite support for now, and consider them as probability
vectors $\vec{p},\vec{q}\in\mathbb{R}^{n}$. To show $q\times\mathrm{Geom}_{1/2}\sqsubseteq p$,
it is equivalent to show that there exist right stochastic matrices
$M_{1},M_{2},\ldots$ with $\{0,1\}$ entries such that $\vec{p}=\sum_{i=1}^{\infty}2^{-i}\vec{q}M_{i}$.
By Lemma \ref{lem:alias}, we have $\vec{p}=\vec{q}M$ for a stochastic
matrix $M$ where each row has at most one positive off-diagonal entry.
For row $x$ with off-diagonal entry $M_{x,a_{x}}=r_{x}/q(x)$, consider
the binary representation of $r_{x}/q(x)$, and put a ``1'' at the
position $(x,a_{x})$ of $M_{j}$ if the $j$-th digit after the decimal
point of the binary representation is ``1'' (otherwise put a ``1''
at the position $(x,x)$) for $j=1,2,\ldots$. This ensures that $M=\sum_{i=1}^{\infty}2^{-i}M_{i}$,
and hence the requirement is satisfied. The following is the complete
proof for the case where the support size may be infinite.
\begin{IEEEproof}
Without loss of generality, assume $p,q$ are pmf's over $\mathbb{N}$
with $p(1)\ge p(2)\ge\cdots$ and $q(1)\ge q(2)\ge\cdots$. Define
pmf $q_{l}$ by
\[
q_{l}(x)=\begin{cases}
q(x) & \mathrm{if}\;x<l\\
\sum_{y=l}^{\infty}q(y) & \mathrm{if}\;x=l\\
0 & \mathrm{if}\;x>l.
\end{cases}
\]
Define $p_{l}$ similarly. Since $p\sqsubseteq p_{l}$, we have $q\preceq p\preceq p_{l}$.
Fix $l$ and let $n>l$ be large enough that $\sum_{y=n}^{\infty}q(y)\le q(l)$,
and hence the $l$ largest entries of $q_{n}$ are the same as the
$l$ largest entries of $q$. Since $p_{l}$ has at most $l$ nonzero
entries, whether $q\preceq p_{l}$ holds only depend on the $l$ largest
entries of $q$. Hence, we have $q_{n}\preceq p_{l}$. By Lemma \ref{lem:alias},
there exists $a_{x}\in[n]\backslash\{x\}$ (we no longer have $a_{x}<x$
since we have to sort $q_{n}(x)$ in descending order before applying
the lemma) and $r_{x}\in[0,q_{n}(x)]$ for $x=1,\ldots,n$ such that
\[
p_{l}(x)=q_{n}(x)-r_{x}+\sum_{y\in[n]:\,a_{y}=x}r_{y}
\]
for any $x=1,\ldots,n$. Define a mapping $g:\mathrm{supp}(q_{n})\times\mathbb{N}\to[n]$
by
\[
g(x,i)=\begin{cases}
x & \mathrm{if}\;2^{i}r_{x}/q_{n}(x)\;\mathrm{mod}\,2<1\\
a_{x} & \mathrm{if}\;2^{i}r_{x}/q_{n}(x)\;\mathrm{mod}\,2\ge1,
\end{cases}
\]
where $a\,\mathrm{mod}\,b:=a-b\lfloor a/b\rfloor$. Since $\sum_{i=1}^{\infty}2^{-i}\mathbf{1}\{2^{i}r_{x}/q_{n}(x)\;\mathrm{mod}\,2\ge1\}=r_{x}/q_{n}(x)$
is the binary representation of $r_{x}/q_{n}(x)$, we have $\mathbf{P}(g(x,Z)=a_{x})=r_{x}/q_{n}(x)$
and $\mathbf{P}(g(x,Z)=x)=1-r_{x}/q_{n}(x)$, where $Z\sim\mathrm{Geom}_{1/2}$.
Let $X\sim q_{n}$ independent of $Z$, we have
\begin{align*}
 & \mathbf{P}(g(X,Z)=x)\\
 & =q_{n}(x)\mathbf{P}(g(x,Z)=x)+\sum_{y\in[n]:\,a_{y}=x}q_{n}(y)\mathbf{P}(g(y,Z)=a_{y})\\
 & =q_{n}(x)-r_{x}+\sum_{y\in[n]:\,a_{y}=x}r_{y}\\
 & =p_{l}(x),
\end{align*}
and hence $q_{n}\times\mathrm{Geom}_{1/2}\sqsubseteq p_{l}$. Since
$q\sqsubseteq q_{n}$, we have $q\times\mathrm{Geom}_{1/2}\sqsubseteq q_{n}\times\mathrm{Geom}_{1/2}\sqsubseteq p_{l}$
by Proposition \ref{prop:kernel_maj}. Since $p_{l}(x)\to p(x)$ as
$l\to\infty$ for any $x\in\mathbb{N}$, by Proposition \ref{prop:agg_closed},
we have $q\times\mathrm{Geom}_{1/2}\sqsubseteq p$.
\end{IEEEproof}
\[
\]

Note that $(\bigwedge S)\times\mathrm{Geom}_{1/2}$ in Theorem \ref{thm:geomsplit}
has an infinite support size or cardinality. If $S$ is finite and
the pmf's in $S$ are over a set $\mathcal{X}$ which is finite, then
we can reduce the cardinality to $|S|(|\mathcal{X}|-1)+1$ (without
increasing its R\'enyi entropy), as given in the following theorem.
\begin{thm}
\label{thm:finite}For a finite set of pmf's $S$ with $|S|=m$, where
the pmf's in $S$ are over a finite set $\mathcal{X}$ with $|\mathcal{X}|=n$,
there exists a pmf $q\in\tilde{\Gamma}(S)$ with $|\mathrm{supp}(q)|\le m(n-1)+1$
and
\[
\Big(\bigwedge S\Big)\times\mathrm{CGeom}_{1/2,m}\preceq q,
\]
where $\mathrm{CGeom}_{1/2,m}$ is the capped geometric distribution
defined in \eqref{eq:cgeom}. As a result, the minimum R\'enyi entropy
of couplings of $S$ satisfies
\[
H_{\alpha}\Big(\bigwedge S\Big)\le H_{\alpha}^{*}(S)\le H_{\alpha}\Big(\bigwedge S\Big)+H_{\alpha}(\mathrm{CGeom}_{1/2,m}).
\]
Note that $H(\mathrm{CGeom}_{1/2,m})=2-2^{2-m}$. Moreover, $q$ and
the aggregation maps for $q\sqsubseteq p$ for all $p\in S$ can be
computed in $O(m^{2}n+mn\log n)$ time (see Algorithm \ref{alg:finitecp}).
\end{thm}
We remark that the cardinality bound $|\mathrm{supp}(q)|\le m(n-1)+1$
is the same as that in \cite{painsky2013memoryless,kocaoglu2017entropic}.
Therefore, the coupling in Theorem \ref{thm:finite} gives a small
R\'enyi entropy, without penalty on the cardinality.

To prove Theorem \ref{thm:finite}, we first show a lemma about coupling
Bernoulli distributions.
\begin{lem}
\label{lem:bern}For a finite set of pmf's $S$ with $|S|=m$, where
the pmf's in $S$ are over $\{0,1\}$, there exists a pmf $q\in\tilde{\Gamma}(S)$
with $|\mathrm{supp}(q)|\le m+1$ and $\mathrm{CGeom}_{1/2,m+1}\preceq q$.
Moreover, $q$ and the aggregation maps for $q\sqsubseteq p$ for
all $p\in S$ can be computed in $O(m^{2})$ time (see Algorithm \ref{alg:bern}).
\end{lem}
\begin{IEEEproof}
We prove the lemma by induction on $m$. The lemma is true when $m=1$
since $\mathrm{CGeom}_{1/2,2}\preceq p$ for any pmf $p$ over $\{0,1\}$.
We now prove the lemma for $m$, assuming that the lemma is true for
any smaller $m$. Let $S=\{p_{1},\ldots,p_{m}\}$. Without loss of
generality, assume $p_{1}$ attains the minimum of $\max\{p_{i}(0),\,p_{i}(1)\}$
for $i=1,\ldots,m$. Let $\gamma:=\max\{p_{1}(0),\,p_{1}(1)\}$. If
$\gamma=1$, all distributions in $S$ are degenerate, and the lemma
clearly holds, and hence we can assume $\gamma<1$. Let $\tilde{p}_{2},\ldots,\tilde{p}_{m}$
be pmf's over $\{0,1\}$ defined as 
\begin{align*}
\tilde{p}_{i}(0) & =\frac{p_{i}(0)-\gamma\mathbf{1}\{p_{i}(0)\ge p_{i}(1)\}}{1-\gamma},\\
\tilde{p}_{i}(1) & =\frac{p_{i}(1)-\gamma\mathbf{1}\{p_{i}(0)<p_{i}(1)\}}{1-\gamma},
\end{align*}
for $i=2,\ldots,m$. Invoke the induction hypothesis to obtain a pmf
$\tilde{q}\in\tilde{\Gamma}(\{\tilde{p}_{2},\ldots,\tilde{p}_{m}\})$
over $[m]$ satisfying $\mathrm{CGeom}_{1/2,m}\preceq\tilde{q}$.
Let $q$ be a pmf over $[m+1]$ with $q(x)=(1-\gamma)\tilde{q}(x)$
for $x\le m$, and $q(m+1)=\gamma$. Since $\gamma\ge1/2$, we have
$\mathrm{CGeom}_{1/2,m+1}\preceq q$. It is left to show that $q\sqsubseteq p_{i}$
for $i=2,\ldots,m$. For $p_{i}$, without loss of generality assume
$p_{i}(0)\ge p_{i}(1)$. Since $\tilde{q}\sqsubseteq\tilde{p_{i}}$,
there exists $A\subseteq[m]$ such that $\tilde{p_{i}}(1)=\sum_{x\in A}\tilde{q}(x)$.
We have 
\[
p_{i}(1)=(1-\gamma)\tilde{p}_{i}(1)=\sum_{x\in A}q(x),
\]
and hence $q\sqsubseteq p_{i}$. Refer to Algorithm \ref{alg:bern}
(which we call the \emph{Bernoulli splitting }algorithm) for the precise
description of the algorithm.
\end{IEEEproof}
\begin{algorithm}[H]
\textbf{$\;\;\;\;$Input:} $\rho_{1},\ldots,\rho_{m}\in[0,1]$ (let
$\rho_{i}=p_{i}(1)$)

\textbf{$\;\;\;\;$Output:} $\{q_{x}\}_{x\in[k]}$, $\{g_{i,x}\}_{i\in[m],x\in[k]}$
(where $k\le m+1$)

\textbf{$\;\;\;\;$$\;\;\;\;$}(let $q_{x}=q(x)$, $g_{i,x}=g_{i}(x)\in\{0,1\}$
for the aggregation mapping for $q\sqsubseteq p_{i}$)

\smallskip{}

\begin{algorithmic}

\State{$g_{i,x}\leftarrow0$ for $i\in[m],x\in[k]$}

\State{$c\leftarrow1$}

\State{$k\leftarrow0$}

\While{$c>0$}

\State{$\gamma\leftarrow\min_{i}\max\{\rho_{i},\,c-\rho_{i}\}$}

\State{$k\leftarrow k+1$}

\State{$q_{k}\leftarrow\gamma$}

\For{$i\leftarrow1,\ldots,m$}

\If{$\rho_{i}\ge c/2$}

\State{$g_{i,k}\leftarrow1$}

\State{$\rho_{i}\leftarrow\rho_{i}-\gamma$}

\EndIf

\EndFor

\State{$c\leftarrow c-\gamma$}

\EndWhile

\Return{$\{q_{x}\}$, $\{g_{i,x}\}$}

\end{algorithmic}

\caption{\label{alg:bern}$\textsc{BernoulliSplitting}(\rho_{1},\ldots,\rho_{m})$}
\end{algorithm}

Algorithm \ref{alg:bern} has time complexity $O(m^{2})$, since after
each iteration of the while loop, the number of $i$'s where $\rho_{i}\in\{0,c\}$
increases by at least one (letting $i^{*}=\mathrm{argmin}_{i}\max\{\rho_{i},\,c-\rho_{i}\}$,
then $\rho_{i^{*}}\in\{0,c\}$ after the iteration), and hence the
number of iterations of the while loop is upper bounded by $m+1$. 

\begin{figure}
\begin{centering}
\includegraphics[scale=0.85]{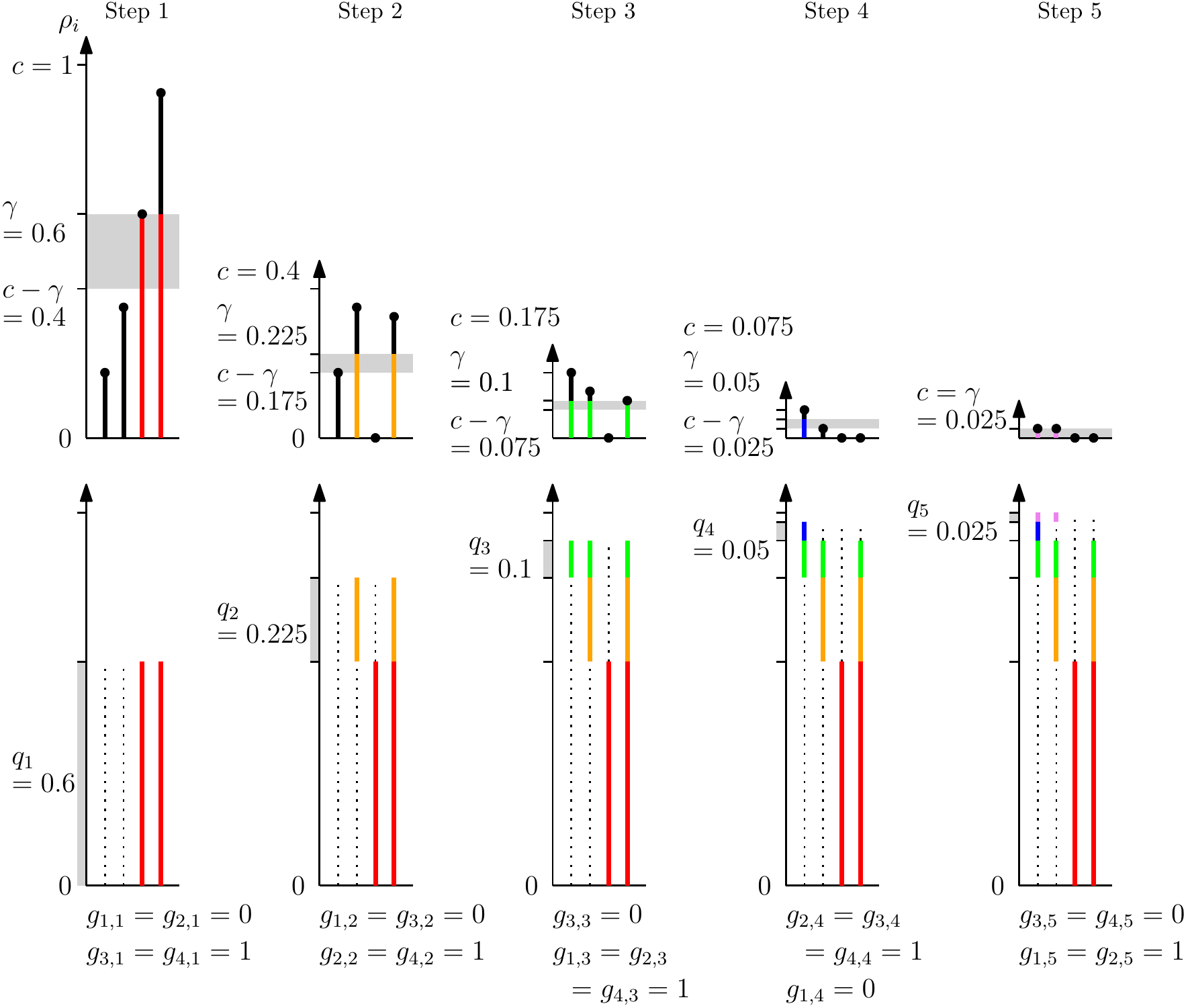}
\par\end{centering}
\caption{\label{fig:bern}Algorithm \ref{alg:bern} applied on $\{\rho_{i}\}_{i}=(0.175,0.35,0.6,0.925)$.}

\end{figure}

Figure \ref{fig:bern} shows Algorithm \ref{alg:bern} applied on
$\{\rho_{i}\}_{i}=(0.175,0.35,0.6,0.925)$. The graphs on top are
$\rho_{i}$ at each iteration of the while loop, and the graphs on
the bottom show $q_{x}$ and $g_{i,x}$ at each iteration. We can
consider the problem as finding a set of sticks with lengths $q_{x}$
which sum to $c$ (initially $c=1$), such that every $\rho_{i}$
(we require $0\le\rho_{i}\le c$) is the sum of the lengths of a subset
of sticks. We use the following greedy approach. If the length of
the longest stick is $\gamma$, then every $\rho_{i}$ must satisfy
either $\rho_{i}\ge\gamma$ if we use the stick to form $\rho_{i}$,
or $\rho_{i}\le c-\gamma$ if we do not use the stick to form $\rho_{i}$
(the regions of inadmissible $\rho_{i}$ are shaded in gray on the
graphs on top). Therefore, the longest possible length of the longest
stick is $q_{1}=\gamma=\min_{i}\max\{\rho_{i},\,c-\rho_{i}\}$. For
every $\rho_{i}$ where $\rho_{i}\ge\gamma$ (or equivalently $\rho_{i}\ge c/2$),
we set $g_{i,1}=1$, meaning that we use the first stick to form $\rho_{i}$,
and then set $\rho_{i}\leftarrow\rho_{i}-\gamma$ (the remainder of
the length to be fulfilled by other sticks). We set $g_{i,1}=0$ for
the rest of $\rho_{i}$. Now the remaining total length of sticks
become $c\leftarrow c-\gamma$. Repeat this process until $\rho_{i}=0$
for all $i$.

\begin{rem}
\label{rem:error}Note that $\sum_{x>L}q(x)\le2^{-L}$ in Lemma \ref{lem:bern}
and Algorithm \ref{alg:bern}. Therefore, stopping Algorithm \ref{alg:bern}
after $L$ steps reduces the time complexity to $O(mL)$, and incurs
an error (in total variation distance) upper bounded by $2^{-L}$,
i.e., it computes a coupling of $\{\mathrm{Bern}(\tilde{\rho}_{i})\}_{i\in[m]}$
instead of $\{\mathrm{Bern}(\rho_{i})\}_{i\in[m]}$, where $|\tilde{\rho}_{i}-\rho_{i}|\le2^{-L}$.
One can also replace ``while $c>0$'' in Algorithm \ref{alg:bern}
to ``while $c>\epsilon$'' (and adjust $q_{x}$ so they sum to 1)
to set the desired error level.
\end{rem}
We now prove Theorem \ref{thm:finite}.
\begin{IEEEproof}
Assume $p_{1},\ldots,p_{m}$ are pmf's over $[n]$ with $p_{i}(1)\ge\cdots\ge p_{i}(n)$.
Let $S=\{p_{1},\ldots,p_{m}\}$, $\bar{q}:=\bigwedge S$. By Lemma
\ref{lem:alias}, let $a_{i,x}\in[x-1]$ and $r_{i,x}\in[0,\bar{q}(x)]$
for $x=2,\ldots,n$, $i=1,\ldots,m$ such that 
\begin{equation}
p_{i}(x)=\bar{q}(x)-r_{i,x}+\sum_{x':\,a_{i,x'}=x}r_{i,x'}\label{eq:alias_rec}
\end{equation}
for any $x=1,\ldots,n$, $i=1,\ldots,m$ (let $r_{i,1}=0$).

Fix any $2\le x\le n$. Since 
\[
\sum_{z=1}^{x-1}\bar{q}(z)=\min_{i\in[m]}\sum_{z=1}^{x-1}p_{i}(z),
\]
there exists $j$ such that $\sum_{z=1}^{x-1}\bar{q}(z)=\sum_{z=1}^{x-1}p_{j}(z)$,
and hence by \eqref{eq:alias_rec} and $a_{j,x}<x$,
\begin{align*}
0 & =\sum_{z=1}^{x-1}\left(p_{j}(z)-\bar{q}(z)\right)\\
 & =\sum_{z=1}^{x-1}\left(\bar{q}(z)-r_{j,z}+\sum_{x':\,a_{j,x'}=z}r_{j,x'}-\bar{q}(z)\right)\\
 & =-\sum_{z=1}^{x-1}r_{j,z}+\sum_{z=1}^{x-1}\sum_{x':\,a_{j,x'}=z}r_{j,x'}\\
 & =\sum_{z=1}^{x-1}\sum_{x'\ge x:\,a_{j,x'}=z}r_{j,x'}\\
 & \ge r_{j,x},
\end{align*}
which means there exists $j$ such that $r_{j,x}=0$. Applying Lemma
\ref{lem:bern} on $\mathrm{Bern}_{r_{i,x}/\bar{q}(x)}$ for $i\neq j$,
let $\tilde{q}_{x}\in\tilde{\Gamma}(\{\mathrm{Bern}_{r_{i,x}/\bar{q}(x)}\}_{i\in[m]\backslash\{j\}})$
be a pmf over $[m]$ with $\mathrm{CGeom}_{1/2,m}\preceq\tilde{q}_{x}$.
We have $\tilde{q}_{x}\sqsubseteq\mathrm{Bern}_{r_{i,x}/\bar{q}(x)}$
for any $i\in[m]$ (this trivially holds when $i=j$).

For $x=1$, let $\tilde{q}_{1}(1)=1$. Since $r_{i,1}=0$, we have
$\tilde{q}_{1}\in\tilde{\Gamma}(\{\mathrm{Bern}_{r_{i,1}/\bar{q}(1)}\}_{i\in[m]\backslash\{j\}})$
and $\mathrm{CGeom}_{1/2,m}\preceq\tilde{q}_{1}$.

Let $q$ be a pmf over $[n]\times[m]$ defined by $q(x,y):=\bar{q}(x)\tilde{q}_{x}(y)$.
For any $i\in[m]$, since $\tilde{q}_{x}\sqsubseteq\mathrm{Bern}_{r_{i,x}/\bar{q}(x)}$,
there exists $A_{x}\subseteq[m]$ such that $\sum_{y\in A_{x}}\tilde{q}_{x}(y)=r_{i,x}/\bar{q}(x)$.
Hence,
\begin{align*}
p_{i}(x) & =\bar{q}(x)-r_{i,x}+\sum_{x'\in[n]:\,a_{i,x'}=x}r_{i,x'}\\
 & =\bar{q}(x)-\bar{q}(x)\sum_{y\in A_{x}}\tilde{q}_{x}(y)+\sum_{x'\in[n]:\,a_{i,x'}=x}\bar{q}(x')\sum_{y\in A_{x'}}\tilde{q}_{x'}(y)\\
 & =\sum_{y\in[m+1]\backslash A_{x}}q(x,y)+\sum_{x'\in[n]:\,a_{i,x'}=x}\sum_{y\in A_{x'}}q(x',y),
\end{align*}
and thus 
\[
g_{i}(x,y):=\begin{cases}
x & \mathrm{if}\;y\notin A_{x}\\
a_{i,x} & \mathrm{if}\;y\in A_{x}
\end{cases}
\]
is an aggregation map for $q\sqsubseteq p_{i}$. Since $|\mathrm{supp}(\tilde{q}_{1})|=1$,
we have $|\mathrm{supp}(q)|\le m(n-1)+1$. We have $\bar{q}\times\mathrm{CGeom}_{1/2,m}\preceq q$
by Proposition \ref{prop:kernel_maj}. Refer to Algorithm \ref{alg:finitecp}
for the precise description of the algorithm.
\end{IEEEproof}
\begin{algorithm}[H]
\textbf{$\;\;\;\;$Input:} pmf's $p_{1},\ldots,p_{m}$ over $[n]$
with $p_{i}(1)\ge\cdots\ge p_{i}(n)$

\textbf{$\;\;\;\;$Output:} $\{q_{x}\}_{x\in[k]}$, $\{g_{i,x}\}_{i\in[m],x\in[k]}$
(where $k\le m(n-1)+1$)

\textbf{$\;\;\;\;$$\;\;\;\;$}(let $q_{x}=q(x)$ for pmf $q$ over
$[k]$,

\textbf{$\;\;\;\;$$\;\;\;\;$ }$g_{i,x}=g_{i}(x)\in[n]$ for the
aggregation mapping for $q\sqsubseteq p_{i}$)

\smallskip{}

\begin{algorithmic}

\State{$\bar{q}\leftarrow\bigwedge_{i\in[m]}p_{i}$}

\For{$i\leftarrow1,\ldots,m$}

\State{$\{a_{i,x}\}_{x=2,\ldots,n},\{r_{i,x}\}_{x=2,\ldots,n}\leftarrow\textsc{MajorizedAlias}(p_{i},\bar{q})$}

\State{$r_{i,1}\leftarrow0$}

\EndFor

\State{$k\leftarrow0$}

\For{$x\leftarrow1,\ldots,n$}

\State{$\{\tilde{q}_{y}\}_{y\in[\tilde{k}]},\{\tilde{g}_{i,y}\}_{i\in[m],y\in[\tilde{k}]}\leftarrow\textsc{BernoulliSplitting}(\{r_{i,x}/\bar{q}(x)\}_{i\in[m]})$}

\State{$q_{k+y}\leftarrow\bar{q}(x)\tilde{q}_{y}$ for $y\in[\tilde{k}]$}

\State{$g_{i,k+y}\leftarrow\mathbf{1}\{\tilde{g}_{i,y}=0\}x+\mathbf{1}\{\tilde{g}_{i,y}=1\}a_{i,x}$
for $i\in[m]$, $y\in[\tilde{k}]$}

\State{$k\leftarrow k+\tilde{k}$}

\EndFor

\Return{$\{q_{x}\},\{g_{i,x}\}$}

\end{algorithmic}

\caption{\label{alg:finitecp}$\textsc{ComputeCoupling}(p_{1},\ldots,p_{m})$}
\end{algorithm}

\medskip{}

\begin{rem}
\label{rem:error_cp}If we perform the modification in Remark \ref{rem:error}
(stopping Algorithm \ref{alg:bern} after $L$ steps), it would reduce
the time complexity of Algorithm \ref{alg:finitecp} to $O(mnL+mn\log n)$,
the support bound to $|\mathrm{supp}(q)|\le(L+1)(n-1)+1$, but incur
an error (in total variation distance on each $p\in S$) upper bounded
by $2^{-L}$, i.e., it computes a coupling of $\{\tilde{p}_{i}\}_{i\in[n]}$
instead of $\{p_{i}\}_{i\in[n]}$, where $d_{\mathrm{TV}}(p_{i},\tilde{p}_{i})\le2^{-L}$.
In practical implementations, setting $L\approx60$ will make the
error negligible compared to floating-point error. Therefore, the
practical running time complexity of Algorithm \ref{alg:finitecp}
is close to $O(mn\log n)$.
\end{rem}

\medskip{}

\section{Acknowledgement}

The author acknowledges support from the Direct Grant for Research,
The Chinese University of Hong Kong. The author would like to thank
the anonymous reviewers for their insightful remarks. In particular,
the author thanks an anonymous reviewer for the suggestion to consider
expressing Lemma \ref{lem:alias} in matrix form.

\medskip{}

\[
\]

\bibliographystyle{IEEEtran}
\bibliography{ref}

\end{document}